\documentclass[aps,prl,twocolumn,superscriptaddress]{revtex4-2}
\usepackage{graphicx}
\usepackage{dcolumn}
\usepackage{bm}
\usepackage{subfigure}
\usepackage{mathrsfs}
\usepackage{multirow}
\usepackage{amsmath}
\usepackage{amsfonts}
\usepackage[makeroom]{cancel}
\usepackage{bbold}
\usepackage[normalem]{ulem}

\usepackage[colorlinks,
linkcolor={red!60!black!},
anchorcolor={green!80!black!},
urlcolor={blue!80!black!},
citecolor={blue!50!black!}]{hyperref}
\usepackage[table]{xcolor}
\usepackage{tipa}
\usepackage{tikz}

\tikzset{
  solidprop/.style={draw, thick}, 
  dottedprop/.style={draw, dashed, thick}
}

\usepackage{array}
\newcolumntype{C}[1]{>{\centering\arraybackslash}m{#1}}
\usepackage{tabularray}

\newcommand{\iu}{\mathrm{i}}

\usetikzlibrary{positioning,arrows}
\usetikzlibrary{decorations.pathmorphing}
\usetikzlibrary{decorations.markings}
\usetikzlibrary{calc}
\usetikzlibrary{calc,angles,quotes,positioning,datavisualization, patterns}
\tikzstyle arrowstyle=[scale=1]

\newif\ifstartcompletesineup
\newif\ifendcompletesineup
\pgfkeys{
	/pgf/decoration/.cd,
	start up/.is if=startcompletesineup,
	start up=true,
	start up/.default=true,
	start down/.style={/pgf/decoration/start up=false},
	end up/.is if=endcompletesineup,
	end up=true,
	end up/.default=true,
	end down/.style={/pgf/decoration/end up=false}
}
\pgfdeclaredecoration{complete sines}{initial}
{
	\state{initial}[
	width=+0pt,
	next state=upsine,
	persistent precomputation={
		\ifstartcompletesineup
		\pgfkeys{/pgf/decoration automaton/next state=upsine}
		\ifendcompletesineup
		\pgfmathsetmacro\matchinglength{
			0.5*\pgfdecoratedinputsegmentlength / (ceil(0.5* \pgfdecoratedinputsegmentlength / \pgfdecorationsegmentlength) )
		}
		\else
		\pgfmathsetmacro\matchinglength{
			0.5 * \pgfdecoratedinputsegmentlength / (ceil(0.5 * \pgfdecoratedinputsegmentlength / \pgfdecorationsegmentlength ) - 0.499)
		}
		\fi
		\else
		\pgfkeys{/pgf/decoration automaton/next state=downsine}
		\ifendcompletesineup
		\pgfmathsetmacro\matchinglength{
			0.5* \pgfdecoratedinputsegmentlength / (ceil(0.5 * \pgfdecoratedinputsegmentlength / \pgfdecorationsegmentlength ) - 0.4999)
		}
		\else
		\pgfmathsetmacro\matchinglength{
			0.5 * \pgfdecoratedinputsegmentlength / (ceil(0.5 * \pgfdecoratedinputsegmentlength / \pgfdecorationsegmentlength ) )
		}
		\fi
		\fi
		\setlength{\pgfdecorationsegmentlength}{\matchinglength pt}
	}] {}
	\state{downsine}[width=\pgfdecorationsegmentlength,next state=upsine]{
		\pgfpathsine{\pgfpoint{0.5\pgfdecorationsegmentlength}{0.5\pgfdecorationsegmentamplitude}}
		\pgfpathcosine{\pgfpoint{0.5\pgfdecorationsegmentlength}{-0.5\pgfdecorationsegmentamplitude}}
	}
	\state{upsine}[width=\pgfdecorationsegmentlength,next state=downsine]{
		\pgfpathsine{\pgfpoint{0.5\pgfdecorationsegmentlength}{-0.5\pgfdecorationsegmentamplitude}}
		\pgfpathcosine{\pgfpoint{0.5\pgfdecorationsegmentlength}{0.5\pgfdecorationsegmentamplitude}}
	}
	\state{final}{}
}
\tikzset{
	fermion/.style={thick,draw=black, postaction={decorate},
		decoration={markings,mark=at position .55 with {\arrow[black]{stealth}}}},
	fermionTP/.style={thick,draw=black, postaction={decorate},
		decoration={markings,mark=at position .27 with {\arrow[black,arrowstyle]{stealth}}}},
	fermionOut/.style={thick,draw=black, postaction={decorate},
		decoration={markings,mark=at position .7 with {\arrow[black,arrowstyle]{stealth}}}},
	photon/.style={decorate, draw=black,
		decoration={coil,aspect=0}},
	ghost/.style={dashed,draw=black,postaction={decorate},
		decoration={markings,mark=at position .55 with{\arrow[black]{stealth}}}},
	graviton/.style={double,thick,draw=black},
	scalar/.style={thick,dashed,draw=black},
	photon1/.style={thick,
		draw=black,decorate,
		decoration={snake, segment length=2mm, amplitude=1mm,post length=.5mm}
	},
photon2/.style={thick,
	draw=black,decorate,
	decoration={snake,pre length=.mm, segment length=2mm, amplitude=1mm,}},
photon3/.style={thick,
	decoration={
		complete sines,
		segment length=1mm,
		amplitude=2mm,
		mirror,
		start up,
		end up
	},
	decorate,
	thick
}
}

\tikzset{
	cross/.pic = {
		\draw[rotate = 45] (-#1,0) -- (#1,0);
		\draw[rotate = 45] (0,-#1) -- (0, #1);
	}
}

\def\centerarc[#1](#2)(#3:#4:#5)
{ \draw[#1] ($(#2)+({#5*cos(#3)},{#5*sin(#3)})$) arc (#3:#4:#5); }
\newcommand{\vertic}{0.5}
\newcommand{\horiz}{0.5}

\tikzset{
	hatched circle/.style={
		pattern=north east lines,
		pattern color=black,
		draw,
		circle,
		minimum size=6pt, 
		inner sep=0pt
	}
}

\begin{document}

\title{Non-Perturbative $S$-matrix Renormalization}

\author{Laurent Freidel} 
\email{lfreidel@pitp.ca}

\affiliation{Perimeter Institute, 31 Caroline Street North, Waterloo, ON, N2L 2Y5, Canada
}

\author{José Padua-Arg\"uelles} 
\email{jpaduaarguelles@pitp.ca}

\affiliation{Perimeter Institute, 31 Caroline Street North, Waterloo, ON, N2L 2Y5, Canada
}
\affiliation{Department of Physics and  Astronomy, University of Waterloo, 200 University Avenue West, Waterloo, ON, N2L 3G1, Canada}

\author{Susanne Schander} 
\email{sschander@pitp.ca}

\affiliation{Perimeter Institute, 31 Caroline Street North, Waterloo, ON, N2L 2Y5, Canada
}

\author{Marc Schiffer} 
\email{marc.schiffer@ru.nl}
\affiliation{High Energy Physics Department, Institute for Mathematics, Astrophysics, and Particle Physics, Radboud University, Nijmegen, The Netherlands}

\begin{abstract}

We propose a renormalization group flow equation for a functional that generates $S$-matrix elements and which captures similarities to the well-known Wetterich and Polchinski equations. While the latter ones respectively involve the effective action and Schwinger functional, which are genuine off-shell objects, the presented flow equation has the advantage of working more directly with observables, i.e. scattering amplitudes. Compared to the Wetterich equation, our flow equation also greatly simplifies the notion of going on-shell, in the sense of satisfying the quantum equations of motion. In addition, unlike the Wetterich equation, it is polynomial and does not require a Hessian inversion. The approach is a promising direction for non-perturbative quantum field theories, allowing one to work more directly with scattering amplitudes.
\end{abstract}

\maketitle

\emph{Introduction.} The $S$-matrix is a fundamental tool in field theory. It is designed to produce physical observables, is related to scattering amplitudes and can be shown to be independent of local field redefinitions \cite{KAMEFUCHI1961529,CHISHOLM1961469,Arzt:1993gz}. 
Understanding how to directly renormalize this quantity using ideas from the Functional Renormalization Group (FRG) is an open issue. Its resolution could give us better control over field redefinitions and allow us to directly renormalize physical observables. It also opens the possibility to better understand the interplay between unitarity and renormalization.\footnote{For early reference on the $S$-matrix and renormalization in string theory see \cite{Banks:1987qs,Hughes:1988bw}.} 

Functional techniques are a fundamental tool when studying Quantum Field Theories (QFTs). Combining such methods with Wilson's Renormalization Group idea gives rise to the FRG, allowing to probe theories within and crucially also beyond their perturbative regime. FRG equations, such as the Polchinski equation \cite{Polchinski-equation} or the Wetterich equation \cite{Wetterich:1992yh, Morris:1993qb, Ellwanger:1993mw}, are based on the Schwinger functional or effective action, respectively. They have been successfully employed, for example, in statistical physics, particle physics, and quantum gravity, see \cite{Dupuis:2020fhh} for a complete review.

Let us now focus on the Wetterich equation, which is based on computing the 1PI effective action including all quantum fluctuations of the theory. While the effective action contains all physical information, the effective action itself is not a physical quantity. Only when going on-shell, i.e, when evaluating it on a solution of the quantum equations of motion, the effective action is directly related to physical quantities like scattering amplitudes. The step of going on-shell is a-priori highly non trivial.   In contrast, the $S$-matrix, which contains all possible scattering amplitudes as its elements, is physical, since scattering amplitudes can be directly measured.

In standard FRG computations, one typically renormalizes the theory based on the effective action, by deriving correlation functions that have integrated out all quantum fluctuations. One then combines those correlation functions into scattering amplitudes see, e.g., \cite{Fu:2022uow, Fu:2024ysj, Ihssen:2024miv, Fu:2025hcm} and \cite{Draper:2020bop, Draper:2020knh, Pastor-Gutierrez:2024sbt} for examples in particle physics and quantum gravity. The last part involves going on-shell, i.e. solving the quantum equations of motion.

In this paper, we propose to renormalize the theory directly on the level of the $S$-matrix, by making use of the generating functional for $S$-matrix elements, focusing on a single scalar field for simplicity. We view this as a complementary approach to standard FRG equations, where other observables, like scaling exponents, are more directly accessible. Combining different FRG equations to study the same physical system might give easier computational access to a more complete set of observables.
To this end, we start by reviewing some properties of such a generating functional, which is structurally similar to the effective action $\Gamma$, in the following section. In the subsequent section titled ``Renormalization flow equation for the S-matrix functional'', we will then apply the standard procedure of deriving the Wetterich equation to the generating functional of $S$-matrix elements. At this point we will also switch to Euclidean signature, which is commonly done in FRG computations, see \cite{PhysRevLett.130.081501} and references therein for exceptions. Interestingly, the resulting equation is equivalent to the Wetterich equation, but has the advantage that `going on-shell' is technically simpler as it only involves the classical kinetic term. Additionally, the equation is polynomial which eliminates the need to invert any Hessian matrix. In the discussion section, we will conclude with an outlook on applications and further developments.\\

\emph{The $S$-matrix Functional.} The starting point for the subsequent $S$-matrix renormalization equation is the $S$-matrix functional \cite{aref1974generating,Jevicki:1987ax}
\begin{equation}
    \mathbb S[\varphi]:=e^{\frac12\iu\int_{M}\mathrm d^D x\varphi(x)K\varphi(x)}Z[J=K\varphi],
    \label{eq:S_definition}
\end{equation}
where $\varphi$ is a scalar field, $D$ the spacetime dimension,
\begin{equation}
    K:=-\square+m^2
    \label{eq:K_definition}
\end{equation}
the physical kinetic operator containing the physical (pole) mass $m$, and $Z$ is the generating functional of correlation functions.

$Z$ is given by (for simplicity, hereafter we omit $\mathrm{d}^D x$ in integrals unless needed.)
\begin{equation}
    Z[J]=\int \mathcal{D}\phi\, e^{\iu I[\phi]+\iu \int J(x)\phi(x)}
    \label{eq:Z_path_int}.
\end{equation}
Here, the path integral is performed over all fields living on a spacetime $\mathcal M$ bounded by times $t = \pm T$ for which $T$ is taken to infinity, and similarly for its pre-factor in \eqref{eq:S_definition}. The classical action 
\begin{equation}
    I[\phi]=\int_{\mathcal M}\left(\frac12\phi K_0\phi-V(\phi)-\frac12\partial^\mu\left(\phi\partial_\mu\phi\right)\right)
\end{equation}
in $Z[J]$ is of second order in $\phi$ which requires the presence of a boundary term given by the latter total derivative contribution.\footnote{The boundary term ensures that on-shell variation of the action are compatible with Dirichlet boundary conditions.
} The bare kinetic operator $K_0$ depends on the bare mass $m_0$.

The functional $\mathbb{S}[\varphi]$ generates $S$-matrix elements according to
\begin{widetext}
\begin{equation} 
    \langle k_1,\dots,k_N|\hat S|p_1,\dots, p_M\rangle=\frac1{\mathbb S[\bar\varphi]}\left(\prod_{j=1}^N\int\mathrm d^{D} y_j \bar{u}_j\right)\left( \prod_{i=1}^M\int\mathrm d^{D} x_i u_i\right) \mathbb S^{(M+N)}[\bar\varphi]_{y_1\dots y_N x_1\dots x_M},
    \label{eq:S_matrix_vs_S}
\end{equation}
\end{widetext}
where the $u_i$'s are positive-frequency modes with momentum $p_i$, evaluated at $x_i$ (and the conjugates are negative-frequency modes of momentum $k_j$, and evaluated at $y_j$). $\bar\varphi$ is a field on-shell of the classical free equation of motion  

\begin{align}\label{eq:on_shell_condition}
K\varphi=0.
\end{align}

$\mathbb S^{(M+N)}[\varphi]$ denotes the $(M+N)$-th functional derivative with respect to $\varphi$ and therefore has arguments (`indices') $z_1,\dots,z_{M+N}$.
\footnote{More precisely for a functional $F[\varphi]$ we adopt the `tensorial' notation $F^{(L)}[\varphi]$ to encode all `components'
$$F^{(L)}[\varphi]_{z_1\dots z_L}:=\frac{\delta}{\delta\varphi(z_1)}\cdots\frac{\delta}{\delta\varphi(z_L)}F[\varphi].$$
As is customary, for small $L$, we may occasionally denote the number of derivatives with a corresponding number of primes. So, for example, $F^{(2)}[\varphi]=F''[\varphi]$. 
\\With this notation, index contractions correspond to spacetime integrals. E.g.
$$\mathrm{Tr}\left(F''[\varphi_0]\right)=\int\frac{\delta}{\delta\varphi(z)}\frac{\delta}{\delta\varphi(z)}F[\varphi]\Bigg|_{\varphi=\varphi_0}.$$
} Taking $N$ of these to be at the future boundary, and $M$ in the past one, and convolving them with free field solutions, gives the $S$-matrix elements.

For an exhaustive proof of this result, we refer to \cite{Nair}, and to comprehend this relation on a more intuitive level, to the supplemental material.\footnote{We caution that the argument relies on the LSZ formalism, which might break down in the presence of long-range interactions \cite{Lippstreu:2025jit}.} 

Before moving on to the derivation of the flow equation, {let us emphasize a boundary formulation which connects the S-matrix point of view with the AdS/CFT perspective and allows a more direct on-shell renormalization.} The generating functional $Z$ can be promoted to a functional $Z_\beta$ of boundary fields $\beta$. In particular, the integral in equation \eqref{eq:Z_path_int} is then confined to fields with Dirichlet boundary conditions $\varphi|_{\partial \mathcal{M}} =\beta$ captured by $\beta$. The corresponding $S$-matrix boundary functional $S_\beta[\varphi]$ reads
\begin{equation}
    \mathbb S_\beta[\varphi]:=e^{\frac12\iu\int\mathrm d^D x\varphi(x)K\varphi(x)}Z_\beta[J=K\varphi].
    \label{eq:S_boundary_definition}
\end{equation}
In a similar fashion as before, it is possible to define $S$-matrix elements according to \cite{Jain:2023fxc, Kim:2023qbl}
\begin{widetext}
\begin{equation}
    \langle k_1,\dots,k_N|\hat S|p_1,\dots,p_M\rangle=\frac1{Z_\beta[0]}\left(\prod_{j=1}^N\int_{\partial\mathcal M}\mathrm d^{D-1} Y_j \bar{u}_j\right)\left(\prod_{i=1}^M\int_{\partial\mathcal M}\mathrm d^{D-1} X_i u_i \right)Z^{(M+N,0)}_\beta[0]_{Y_1\dots Y_N X_1\dots X_M}\bigg|_{\beta=0},
    \label{eq:S_matrix_vs_Z_Dirichlet}
\end{equation}
\end{widetext}
where the integrals are done over points $X_i$ and $Y_j$ in the spacetime boundary $\partial\mathcal M$ (or more precisely on the boundary of the slab that is taken to infinity), and the functional derivatives of $Z_\beta[J]$ are only taken with respect to $\beta$, hence the superscript $(M+N,0)$.  We  refrain from providing a detailed discussion and derivation of this result. To the best of our knowledge, no method exists to deduce a closed flow equation for $\mathbb{S}_\beta$ when treated as a functional of $\beta$, but we will come back to this point in the discussion. The flow equation derived below, however, is one for a functional $\mathbb{S}_\beta(\varphi)$, which when taking $\varphi$ to be on-shell, i.e. $K\varphi=0$, generates $S$-matrix elements upon taking functional derivatives with respect to $\beta$ — apart from also generating $S$-matrix elements in the sense of \eqref{eq:S_matrix_vs_S}.\\

\emph{Renormalization flow equation for the $S$-matrix functional.} Now we will proceed to derive a renormalization flow equation for the $S$-matrix functional $\mathbb S$, or rather its Wick-rotated version. More precisely, we chose to perform the computations in Euclidean signature, as it provides a more rigorous mathematical framework and allows for a more straightforward comparison with standard FRG results, see, e.g. \cite{Floerchinger:2011sc, Manrique:2011jc, Fehre:2021eob, DAngelo:2022vsh, DAngelo:2023wje, Saueressig:2025ypi} for various paths to extending the Wetterich equation to Lorentzian spacetimes. For notational simplicity, we will use the same symbols for the Wick-rotated quantities as before. We caution that after Wick-rotation, the $S$-matrix elements are not directly related to physical quantities, even after going on-shell — the Wick-rotation back to Lorentzian signature is still required. Explicitly, we consider
\begin{equation}
    \mathbb S[\varphi]:=e^{-\frac12\int \varphi K\varphi}Z[J=K\varphi].
    \label{eq:S_definition_Euclidean}
\end{equation}
The derivation proceeds along the lines of the standard FRG computations, see, e.g., \cite{Wetterich:1992yh,  Morris:1993qb, Ellwanger:1993mw, Pawlowski:2005xe, Gies:2006wv} and starts by introducing a regulator term into the action
\begin{align}
    \Delta I_k[\varphi] := \frac12 \mathrm{Tr}(\varphi R_k \varphi)= \frac12 \int \frac{\mathrm{d}^4 q}{(2\pi)^4} \varphi(-q) R_k(q) \varphi(q).\nonumber
\end{align}
In momentum space the  regulator kernel is a function $R_k(q)$ which should satisfy
\begin{gather}
    \lim_{q^2/k^2 \rightarrow 0} R_k(q) >0,\quad\lim_{k^2/q^2 \rightarrow 0} R_k(q) =0,\nonumber\\
    \text{and}\quad \lim_{k^2 \rightarrow \Lambda \rightarrow \infty} R_k(q)\rightarrow \infty ,
    \label{eq:S_k_definition}
\end{gather}
and $\Lambda$ is a UV-cutoff, cf. \cite{Gies:2006wv}. $\Delta I_k[\varphi]$ represents a momentum-dependent mass term which we add to the standard action functional.

By virtue of the regulator and inspired by the definition of $\mathbb{S}$ in equation \eqref{eq:S_definition}, we define the FRG $S$-matrix functional as
\begin{align} \label{eq:Regulated-S}    \mathbb{S}_k[\varphi] := e^{-\frac12 \int \varphi K_k \varphi} Z_k[J = K_k \varphi]
\end{align}
where
\begin{align}\label{eq:Kk}
    K_k := K + R_k
\end{align}
is the regulated physical kinetic operator and the regulated generating functional of correlation functions is given by
\begin{align}
    Z_k[K_k \varphi] = \int \mathcal{D}\phi\, e^{-I_{k}[\phi] + \int{\phi} K_k \varphi} 
    \label{eq:Z_k}
\end{align}
with 
$ I_{k}[\phi] := I[\phi] + \Delta I_k[\phi].
$
%
With these definitions, we recover the standard $S$-matrix generating functional $\mathbb{S}$ for $k\to0$, where the regulator $R_k$ vanishes, indicating that all quantum fluctuations have been integrated out.

{The goal is to derive the following FRG flow equation for $\mathbb S_k$ \cite{Wetterich:1992yh, Morris:1993qb, Ellwanger:1993mw}:
\begin{align}
    \partial_t \mathbb{S}_k = \frac12 \mathrm{Tr} \left( \partial_t {K_k^{-1}} \left(\mathbb{S}_k'' + \mathbb{S}_k K_k \right) \right).
    \label{eq:flow_equation}
\end{align}
It is analogous to the well-known Wetterich equation for the regularized effective action. While the Wetterich equation is initialized with the classical action, we initialize the S-matrix by the tree level amplitude.}

For this purpose, we start by deriving a flow equation for the functional
\begin{align} \label{eq:T-S-relation}
    \mathbb{T}_k[\varphi] := - \log \mathbb{S}_k [\varphi].
\end{align}
%
%
%
If we define the regulated \emph{free} action by
\begin{align}
    I_{0,k}[\varphi] := \frac12 \int  \varphi K_k \varphi\,,
\end{align}
we see that 
\begin{equation} \label{eq:T-functional}
    \mathbb{T}_k [\varphi] = I_{0,k}[\varphi] - W_k [K_k \varphi]\,,
\end{equation}
where
\begin{align}
    W_k[J] := \log Z_k[J],
\end{align}
is the regulated generating functional of connected correlators.

Eq.~\eqref{eq:T-functional} shows that $\mathbb{T}_k$ is structurally similar to a regulated effective action $\Gamma_k[\Phi]$. Indeed, $\Gamma_k[\Phi]$ is the Legendre transform of $W_k[J]$, changing the source argument $J$ to the vacuum expectation value $\Phi$ of the quantum field $\phi$ in \eqref{eq:Z_k}. Further, \eqref{eq:T-functional} looks like such transform, but setting $\Phi=\varphi$ and $J=K_k\varphi$. Incidentally, we also see that, up to an overall sign, $\mathbb{T}_k$ is obtained by subtracting the free action from $W_k[K_k\Phi]$, which is reminiscent of the Wilsonian effective action, see also \cite{Ihssen:2022xjv}. Consequently, we are able to follow the steps of the popular derivation of the Wetterich equation for the effective action.

Note that equation \eqref{eq:T-functional} also allows us to parse common QFT quantities in terms of the new functionals $\mathbb T_k$ and $\mathbb{S}_k$. This naturally holds also for the unregulated functionals. For example, the field expectation value $\Phi$ at fixed source can be expressed in terms of $\mathbb T_k$ by using the known identity $\Phi[J] = W'[J]$. Indeed, this identity and equation \eqref{eq:T-functional} lead to
\begin{equation}    \label{eq:Phi-def}            \Phi[K_k\varphi]=W_k'[K_k\varphi]=\varphi-{K_k^{-1}} \mathbb{T}_k'[\varphi].
\end{equation}
Now we will derive a renormalization flow equation for the  functional $\mathbb T_k$. Therefore, we consider the customary $\partial_t \mathbb{T}_k[\varphi]$ where $\partial_t = k \,{\tfrac{d}{dk}}$. We have
\begin{align} \label{eq:T-der-t}
    \partial_t \mathbb{T}_k [\varphi] = \partial_t I_{0,k}[\varphi] - \partial_t (W_k [K_k \varphi]).
\end{align}
The first term of the right hand side can at once be identified as $\frac12 \int \varphi \, (\partial_t R_k) \varphi=\frac12(\partial_t\Delta I_k)[\varphi]$, whereas for the second we need to take into account both the implicit and explicit $k$-dependence. Then, the chain rule yields
\begin{align} \label{eq:chain_rule_step}
    \partial_t (W_k [K_k \varphi]) = (\partial_t W_k)[K_k \varphi] + \int W_k'[K_k \varphi] \, \partial_t (K_k \varphi).
\end{align}
A straight-forward calculation (see \cite{Wetterich:1992yh, Pawlowski:2005xe, Gies:2006wv}) shows that the derivative acting on $W_k$ at fixed field argument produces a two-point function, which can in turn be expressed in terms of $W_k''$ by adding the disconnected components to it. We get the Polchinski-type equation
\begin{equation}
\partial_t W_k[J]= -\frac12\mathrm{Tr} \left(\partial_t R_k\left( W_k''- W_k'^{\otimes 2}\right)\right). 
\label{eq:Polchinski}
\end{equation}
Both $W_k''$ and these disconnected components can then be rewritten in terms of $\mathbb T_k$ by using equation \eqref{eq:Phi-def}. Using again equation \eqref{eq:Phi-def} for the second term on the right hand side of equation \eqref{eq:chain_rule_step}, equation \eqref{eq:T-der-t} eventually yields
\begin{align} \label{eq:T_k-t-der-2}
    \partial_t \mathbb{T}_k [\varphi] = \frac{1}{2} \mathrm{Tr} &\left( (\partial_t R_k ) (\mathbb{1} - {K_k^{-1}} \mathbb{T}_k'') {K_k^{-1}} \right)\\\nonumber
    &+(\partial_t \Delta I_k)[{K_k^{-1}} \mathbb T_k'].
\end{align}
Using the equality $\partial_t R_k=\partial_t {K_k}$ that follows from \eqref{eq:Kk}, the first term can be rewritten as $- \frac12 \mathrm{Tr} \left( \partial_t {K_k^{-1}} \left( K_k - \mathbb{T}_k'' \right) \right)$ by using the cyclic property of the trace. Similarly, we can express the second term as a trace by introducing $\mathbb T_k'^{\otimes 2}:=\mathbb T_k'\otimes\mathbb T_k'$. 
Finally, we obtain the flow equation for $\mathbb T_k$
\begin{align} \label{eq:T_k-flow-equation}
    \partial_t \mathbb{T}_k = \frac12 \mathrm{Tr} \left( \partial_t {K_k^{-1}} \left( \mathbb{T}_k'' - \mathbb{T}_k'^{\otimes 2} -K_k \right) \right).
\end{align}
Up to the last term, which is field-independent and thus irrelevant when studying the flow of field derivatives of $\mathbb{T}_k$, this coincides with the well-known Polchinski-type equation \eqref{eq:Polchinski}. {Re-expressing $\mathbb T_k$ in terms of $\mathbb S_k$ one recovers \eqref{eq:flow_equation}.}


It is interesting to introduce the free-field free energy 
\begin{equation}
F_k = \frac12\mathrm{Tr} \ln K_k
\end{equation}
and use it to renormalize the $S$-matrix functional. That is, we define a renormalized $S$-matrix functional 
\begin{align}
 \widetilde{\mathbb{S}}_k 
 := e^{F_k} \mathbb{S}_k
\end{align}
which satisfies the remarkably simple equation 
\begin{align}\label{eq:flow_equation_tilde}
  \partial_t \widetilde{\mathbb{S}}_k = \frac12 \mathrm{Tr} \left( \partial_t {K_k^{-1}} \widetilde{\mathbb{S}}_k'' \right),
\end{align}

It is illustrative to write \eqref{eq:flow_equation} diagrammatically by introducing the notation

\begin{align}
K_k^{-1} 
&= \tikz[baseline=-0.5ex]\draw[dottedprop] (0,0)--(1.4,0);\quad\text{and}\quad
\partial_t K_k=\partial_t R_k
=\tikz[baseline=-0.5ex]{
  \node[
    draw,
    circle,
    preaction={fill=white},
    minimum size=0.3cm,
    inner sep=0pt,
    path picture={\draw (path picture bounding box.south east) --
                        (path picture bounding box.north west)
                        (path picture bounding box.south west) --
                        (path picture bounding box.north east);}
  ] at (0.6,0) {};
}
\end{align}

Then, the flow equation \eqref{eq:flow_equation_tilde} can be represented as
\begin{alignat}{1}
        \partial_t
        \begin{tikzpicture}[baseline=(aux2.base),node distance=\vertic cm and \horiz cm,cross/.style={path picture={ 
					\draw[black]
					(path picture bounding box.south east) -- (path picture bounding box.north west) (path picture bounding box.south west) -- (path picture bounding box.north east);
			}}]
			\coordinate[] (e1);
			\coordinate[right=of e1] (f1);
			\coordinate[above=1.5*\horiz cm of f1] (aux1);
			\coordinate[right=of e1] (e2);
			\coordinate[right= of e2] (h1);
			\coordinate[above=of aux1](aux2);
			\coordinate[above=of aux2](aux3);
			\coordinate[right=of aux2](aux4);
			\coordinate[right=of aux4](aux5);
			\node[
			draw,
			circle,
			fill=gray!40,      
			preaction={fill=white}, 
			minimum size=0.3cm,     
			inner sep=0.5pt           
			] at ([yshift=0.1cm]aux5) {$\widetilde{\mathbb{S}}_k$};  
		\end{tikzpicture}%
        =
			-\frac{1}{2}\,\,
		\begin{tikzpicture}[baseline=(aux2.base),node distance=\vertic cm and \horiz cm,cross/.style={path picture={ 
				\draw[black]
				(path picture bounding box.south east) -- (path picture bounding box.north west) (path picture bounding box.south west) -- (path picture bounding box.north east);
		}}]
		\coordinate[] (e1);
		\coordinate[right=of e1] (f1);
		\coordinate[above=1.5*\horiz cm of f1] (aux1);
		\coordinate[right=of e1] (e2);
		\coordinate[right= of e2] (h1);
		\coordinate[above=of aux1](aux2);
		\coordinate[above=of aux2](aux3);
		\draw[scalar] (aux2) circle (\vertic cm);
		\node[
		draw,
		circle,
		fill=gray!40,      
		preaction={fill=white}, 
		minimum size=0.3cm,     
		inner sep=0.5pt           
		] at ([yshift=0.1cm]aux1) {$\widetilde{\mathbb{S}}_k$};  
		\node[draw,	circle,	cross,	minimum size=0.17cm, 
		inner sep=0pt,preaction={fill=white}] at (aux3){};
		\end{tikzpicture}\quad.
		\end{alignat}

Equation \eqref{eq:flow_equation} (or \eqref{eq:flow_equation_tilde}) is our main result: a renormalization group flow equation analogous to the Wetterich equation but formulated in terms of a QFT functional that can be understood as the Wick-rotated generator of $S$-matrix elements. The two equations are equivalent, which follows from the exact relations between the respective generating functions, see Figure 1 in the supplemental material. However, each gives more direct access to different quantities: For the Wetterich equation these are correlation functions, whereas for \eqref{eq:flow_equation} these are $S$-matrix elements — see Table I in the supplemental material for a more explicit comparison (For the most general renormalization flow equation see \cite{Wegner:1974sla}, and see \cite{Ihssen:2022xjv} for a summary of known flow equations derived thereof.). \\



\emph{Discussion.} The derivation of \eqref{eq:flow_equation} (or \eqref{eq:flow_equation_tilde}) was achieved through the introduction of an auxiliary functional that is structurally similar to the effective action, defined via a transformation that is no longer a Legendre transform.
This means that the knowledge of $\mathbb S_k$ is equivalent off-shell to the knowledge of the quantum effective action. However, a key consequence of this modification is that going on-shell simplifies to solving the free wave equation with the physical mass –in line with the picture of working more directly with on-shell quantities like scattering amplitudes. This stands in contrast to the standard approach, where solving the quantum equations of motion via the effective action can be cumbersome, particularly in the presence of non-trivial boundary conditions or in spacetimes lacking Poincaré symmetry.

This perspective naturally suggests further exploration of the $S$-matrix boundary functional discussed above, which is fundamentally an on-shell object. The first goal in this program would be to derive a flow equation for this boundary functional.  One possible approach is to determine whether equation \eqref{eq:flow_equation} can be reformulated as a boundary flow equation for the boundary functional. However, this is \emph{prima facie} challenging, as projecting the equation onto a shell parametrized by boundary conditions $\beta$ does not necessarily guarantee that the Hessian has vanishing components in the off-shell directions. Nevertheless, whether this can be achieved via eq.~\eqref{eq:flow_equation} through the use of boundary-to-bulk propagators, specific choices of regulators, regularized notions of on-shellness, or leveraging the fact that the problematic term appears inside a trace, are all directions that could be explored in future work. Alternatively, a promising approach is to adapt the framework of \cite{JanDraft}, where an infrared cutoff scheme imposes a finite spacetime volume, leading to a flow equation where the trace is naturally performed over the boundary. Thus, such an approach could provide a more direct path toward a well-defined boundary flow equation.
It would be interesting to connect this to consequences of unitarity such as the Cutkosky cutting rules 
\cite{Cutkosky:1960sp}.

Beyond these qualities, equation \eqref{eq:flow_equation} possesses yet another appealing feature: it is polynomial in form. This stands in contrast to the Wetterich equation, which involves the inverse Hessian of the effective action. If this polynomial structure translates into practical simplifications, it could facilitate the development of more efficient non-perturbative methods.

It is interesting that in our derivation of the renormalization flow for the $S$-matrix functional we encounter the regularized free energy which is the term that contributes to the cosmological constant renormalization.  Recently, it was emphasized by \cite{Becker:2020mjl,Freidel:2022ryr,Ferrero:2024yvw} that the free energy is  a dimensionless entity and as such can be logarithmically regularized leading to a resolution of the cosmological constant problem.

Another key aspect for future investigation is the Wick rotation back to Lorentzian signature, ensuring a direct connection to the $S$-matrix. Alternatively, it would be interesting to see, if recent developments to formulate the Wetterich equation in Lorentzian spacetimes \cite{Floerchinger:2011sc, Manrique:2011jc, Fehre:2021eob, DAngelo:2022vsh, DAngelo:2023wje, Saueressig:2025ypi} can be adapted to the flow equation of $S$-matrix elements.

Taken together, these features suggest a promising direction for non-perturbative QFT computations, allowing one to work directly with scattering amplitudes while circumventing the need to compute correlation functions first. Further exploration is warranted to assess both the advantages and potential limitations of this approach. A natural next step is to apply equation \eqref{eq:flow_equation} in concrete settings, developing truncation schemes tailored to this framework — analogous to those successfully employed in the effective action approach. {A first step would be to recover known 1-loop results for scattering amplitudes. To this end, using tree-level scattering as the initial ansatz for $\mathbb{S}_k$, would be instructive.} This would help assess the utility and limitations of using this flow equation for non-perturbative QFT calculations. Ultimately, employing the Wetterich and $S$-matrix flow equations in a combined fashion, might allow to efficiently compute different sets of observables like scattering amplitudes, scaling exponents, etc.\\

\emph{Acknowledgements.} We thank Francois David, Victoria Knapp-Pérez, Benjamin Knorr, Jan Pawlowski, and Shouryya Ray for discussions and Benjamin Knorr, and Jan Pawlowski for comments on the manuscript. This project originated during the quantum-gravity group retreat organized by Florian Girelli and Marc Schiffer, which created the opportunity for several quantum-gravity researchers to work together in an exceptional environment, on problems that were crossing boundaries. This project would not have seen the light of day without it.
This research is supported by Perimeter Institute.
Research at Perimeter Institute is supported in part by the Government of Canada through the Department of Innovation, Science and Economic Development Canada and by the Province of Ontario through the Ministry of Colleges and Universities. The work of J.~P.~A.~is supported by an NSERC grant awarded to Bianca Dittrich. The work of M.~S.~was in parts supported by a Radboud Excellence fellowship from Radboud University in Nijmegen, Netherlands.

\bibliography{references}

\clearpage
\onecolumngrid
\section{Supplementary material}
\subsection{Generating S-matrix elements}

To gain some understanding of the relation

\begin{equation} 
    \langle k_1,\dots,k_N|\hat S|p_1,\dots, p_M\rangle=\frac1{\mathbb S[\bar\varphi]}\left(\prod_{j=1}^N\int\mathrm d^{D} y_j \bar{u}_j\right)\left( \prod_{i=1}^M\int\mathrm d^{D} x_i u_i\right) \mathbb S^{(M+N)}[\bar\varphi]_{y_1\dots y_N x_1\dots x_M},
    \label{eq:S_matrix_vs_S}
\end{equation}

let us begin by considering the first functional derivative of $\mathbb S$. It is given by
\vspace{-6pt}
\begin{equation}
    \mathbb{S}'[\varphi] = e^{\iu\frac12 \int \varphi K \varphi} \left(\iu
    K \varphi \, Z[K \varphi] + Z'[K \varphi] K  \right).
    \label{eq:S_prime}
\end{equation}

Interestingly, only the second term of $\mathbb{S}'[\varphi]$ contributes to the scattering amplitudes whenever $\varphi$ is on-shell of the classical free equation of motion \eqref{eq:on_shell_condition}.

Equation \eqref{eq:S_prime} allows us to infer that the $n$-th functional derivative of $\mathbb S$ will consist of a combination of terms with $K$, $\varphi$ and functional derivatives of $Z$. Setting $\varphi$ to be any solution to the on-shell condition \eqref{eq:on_shell_condition} will make several of these terms vanish. And similarly, when contracting with the (\emph{physical}) modes as in \eqref{eq:S_matrix_vs_S} other terms will vanish. In fact, for any finite contribution to the contraction of the modes with $\mathbb{S}^{(M+N)}$, the polynomial degree of $K$ must exactly match the number of derivatives of $Z$. This is because only then the zeros from the $K$'s (in momentum space) are counterbalanced by the (exact same number of) poles from the derivatives of $Z$. Recall that $n$-point functions have poles of order $n$ when going to the mass shell. 

In summary, in the context of the contraction with the modes in equation \eqref{eq:S_matrix_vs_S}, we can make the substitution
\begin{align}
    \mathbb{S}^{(n)}[\bar\varphi] \leftrightarrow  Z^{(n)}[\bar\varphi] K^n,
\end{align}
with a proper contraction of the operators in the right hand side.

Therefore, we have

\begin{gather}
    \left(\prod_{j=1}^N\int\mathrm d^{D} y_j \bar{u}_j\right) \left(\prod_{i=1}^M\int\mathrm d^{D} x_i u_i\right) \mathbb S^{(M+N)}[\bar\varphi]_{y_1\dots y_N x_1\dots x_M}\nonumber\\
    =\nonumber\\
    \left(\prod_{j=1}^N\int\mathrm d^{D} y_j \bar{u}_j K_j\right)\left( \prod_{i=1}^M\int\mathrm d^{D} x_i u_i K_i\right) Z^{(M+N)}[J=K\varphi]_{y_1\dots y_N x_1\dots x_M}\bigg|_{\bar\varphi}.
    \label{eq:S_LSZ}
\end{gather}

The lower-part of this equation corresponds to the well-known LSZ prescription, and hence to the anticipated $S$-matrix element, see equation \eqref{eq:S_matrix_vs_S}.

\subsection{Connection between functionals and flow equations}

To visualize the relation between the $S$-matrix generating functional $\mathbb S$ introduced in the main article, the Euclidean partition function $Z$, and the effective action $\Gamma$, we refer to Figure I below. To compare our $\mathbb S$-matrix functional flow equation with the Wetterich and the Polchinski flow equations respectively, please see Table \ref{tab:comparison}.

\begin{figure}[h]
\centering
    \includegraphics[width=0.6\linewidth]{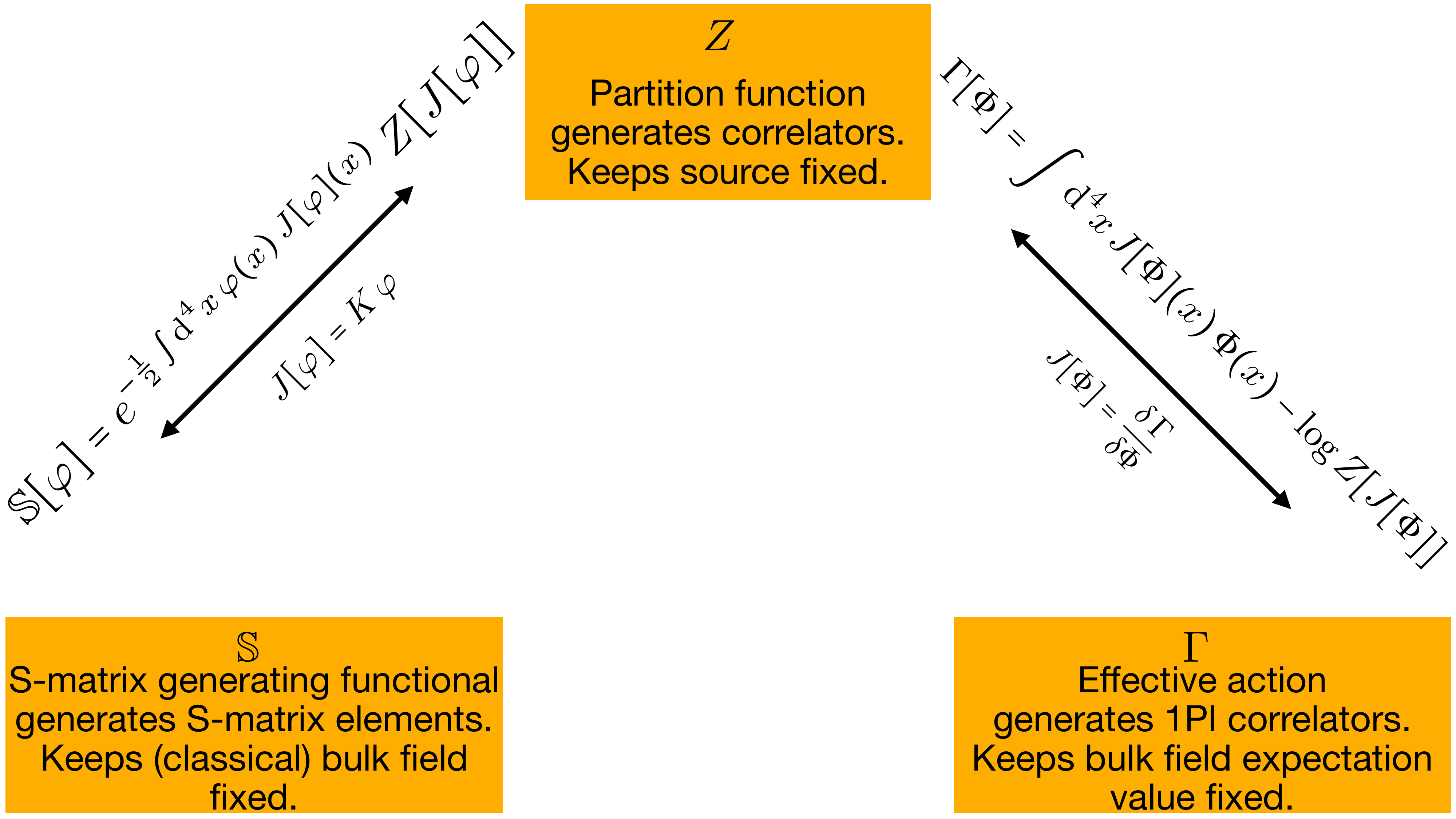}
    \caption{Relations between the (Euclidean) partition function $Z$, effective action $\Gamma$, and $S$-matrix generating functional $\mathbb{S}$. The partition function and the effective action are related via a Legendre transform, while the $S$-matrix generating functional is related to the partition function via $\mathbb S[\varphi]:=e^{-\frac12\int \varphi K\varphi}Z[J=K\varphi]$, Eq. (12) in the main text.}
    \label{fig:triangle}
\end{figure}

\begin{table}[t]
\centering
\begin{tblr}{
  colspec={
    |Q[c,m,6cm]
    |Q[c,m,3.1cm]
    |Q[c,m,3.1cm]
    |Q[c,m,3.1cm]|
  },
  hlines,
  row{1} = {font=\bfseries},
  rowsep = 6pt
}
Equation
&
Generates
&
Complexity
&
Quantum equation of motion
\\

Wetterich:\par\medskip
$\displaystyle
\partial_t\Gamma_k[\Phi]
=
\frac{1}{2}\mathrm{Tr}\left(
\partial_t R_k
\left(\Gamma_k''+R_k\right)^{-1}
\right)
$
&
1PI diagrams
&
Requires inversion\par of Hessian
&
$\displaystyle
\frac{\delta \Gamma}{\delta \Phi}=0
$\par\medskip
Highly non-trivial on curved/dynamical spacetimes or with flexible boundary conditions.
\\

Polchinski:\par\medskip
$\displaystyle
\partial_t W_k[J]
=
-\frac{1}{2}\mathrm{Tr}\left(
\partial_t R_k
\left(W_k''+W_k'^{\otimes 2}\right)
\right)
$
&
Connected diagrams
&
Polynomial
&
$\displaystyle
J=0
$\par\medskip
Trivial: setting argument/source to zero.
\\

Eq.~(27) in main article:\par\medskip
$\displaystyle
\partial_t \mathbb{S}_k[\varphi]
=
\frac{1}{2}\mathrm{Tr}\left(
\partial_t K_k^{-1}
\left(\mathbb{S}_k''+\mathbb{S}_k K_k\right)
\right)
$
&
$S$-matrix\par elements
&
Polynomial
&
$\displaystyle
K\varphi=0
$\par\medskip
Classical free equation of motion---note that $\varphi\equiv0$ satisfies it.
\end{tblr}

\caption{Comparison between the approaches to FRG using the Wetterich, Polchinski and $S$-matrix functional flow equations.}
\label{tab:comparison}
\end{table}

\end{document}